\documentclass[10pt,tightenlines,amsmath,
amssymb,nofootinbib,prd,twocolumn,showpacs,preprintnumbers]{revtex4}

\newcommand{\beq}{\begin{equation}}
\newcommand{\eeq}{\end{equation}}
\newcommand{\bea}{\begin{eqnarray}}
\newcommand{\eea}{\end{eqnarray}}
\newcommand{\nn}{\nonumber}
\newcommand{\junk}[1]{}
\def\<{\langle}
\def\>{\rangle}
\def\d{\partial}
\def\+{\dagger}
\def\UA{$U(1)_{A}~$}
\def\UB{$U(1)_{B}~$}
\def\UY{$U(1)_{Y}~$}
\def\UEM{$U(1)_{EM}~$}
\def\mueff{\mu_{\mathrm{eff}}}

\def\meffk0{m_{\mathrm{eff}}^{0^{2}}}
\def\kzero{K^0}
\def\kplus{K^+}
\def\vpi{v}
\def\fpi{f_{\pi}}
\def\thetak{\theta_{K^0}}

\def\thetacrit{\theta_{\mathrm{crit}}}

\def\trace{\mathrm{Tr}}
\def\diag{\mathrm{diag}}
\def\wt{\tilde{w}}
\def\mdw{m_{\mathrm{dw}}}

\begin{document}

\title{Drum vortons in high density QCD}
\author{Kirk~B.~W.~Buckley, Max~A.~Metlitski, and Ariel~R.~Zhitnitsky}
\affiliation{Department of Physics and Astronomy, \\
University of British Columbia, \\
Vancouver, BC V6T 1Z1, Canada  }
\date{\today}
\begin{abstract}
Recently it was shown that high
density QCD supports of number of topological defects. 
In particular, there are $U(1)_Y$ strings that 
arise due to $K^0$ condensation that occurs when the strange quark 
mass is relatively large. 
The unique feature of these strings is that they possess a nonzero
$K^+$ condensate that is trapped on the core. 
In the following we will show that these strings (with nontrivial 
core structure) can form closed
loops with conserved charge and currents trapped on the string world sheet.
The presence of conserved charges allows
these topological defects, called vortons, to carry angular momentum, 
which makes them classically stable objects. We also give arguments 
demonstrating that vortons carry angular momentum  very 
efficiently (in terms of energy per unit angular  momentum) such 
that they might be the important degrees of freedom in the cores 
of neutron stars.   
\end{abstract}

\maketitle

\section{Introduction}

In the high baryon density regime it is well established
that QCD behaves like a BCS
superconductor where the condensate of electron Cooper pairs
is replaced by a diquark condensate which has a very specific
structure \cite{baillov,colorsc,colorsc2,rapp,CFL} (see \cite{csrev}
for a review of high density QCD). The presence of a condensate of quark
Cooper pairs breaks certain symmetries which are
respected at zero baryon density.
Recently, it has been shown that a consequence of this
symmetry breaking is the appearance of various
topological defects \cite{ssz,sonKaon,fzstrings,krstrings,kirkdw,superk}.
This leads to the exciting possibility that such
objects (which do not exist in the standard model in the vacuum)
may be present within the interior of compact stars
where such large densities may be realized \cite{NASA}.

In addition to a diquark condensate, it may be energetically
favorable for a $\kzero$ condensate to form
for the physical value of the strange quark mass,
$m_s \simeq 150~ \mathrm{MeV} \gg m_u,m_d$ 
\cite{schafer,bedascha,kaplredd,bedaque}.
The hypercharge \UY symmetry is spontaneously broken in this case,
leading to the formation of global \UY strings
\cite{fzstrings}.
The main feature which distinguishes these strings
from other types of global strings present at high density,
namely \UB and \UA strings
\cite{fzstrings}, is the presence of a nonzero $\kplus$
condensate which forms on the core of the string \cite{krstrings,superk}.
This means that the  electromagnetic
\UEM symmetry is spontaneously broken in the core of the string, and 
we will refer to them as superconducting $K$ strings. Analytical
calculations demonstrated explicitly that $\kplus$ condensation only occurs
for a relatively large strength of the kaon condensate, parametrized 
by the angle $\thetak$ (see below). 
If $\theta < \thetacrit \sim \sqrt{(m_d-m_u)/m_s}$, the symmetry is 
restored in the core and $K^+$ condensation does not occur \cite{superk}.
We should note that this phenomenon where the symmetry is not restored
in the core of the vortex 
was pointed out long ago by Witten \cite{witten} and has
received alot of attention since in the context of cosmology
and cosmic strings \cite{hindmarsh,turner,davis,davisshell1}.
It is interesting to note that an example of this phenomenon has been found
in the context of the $SO(5)$ theory of high temperature
superconductivity \cite{highTc}, where the conventional vortices in
type II superconductors might have antiferromagnetic cores. 
Furthermore, there has been recent experimental evidence supporting
the presence of the antiferromagnetic condensate in superconducting vortices
\cite{highTcexp}. Therefore, the study of vortices when the symmetry 
is not restored in the core is not a purely academic question. 

In addition to the work of Witten \cite{witten}, 
the subsequent studies demonstrated that
the presence of a condensate localized on the string may lead to
the classical stability of superconducting string loops
\cite{hindmarsh,turner}. The first class of superconducting string loops,
called ``springs,'' were characterized by a
nonzero spacelike current 
trapped on the string world sheet
\cite{hindmarsh,turner}. It is well known that a loop of a normal global
string with no condensate in the core is unstable in the vacuum.
Upon formation the loop will shrink and eventually disappear
through the emission of particles.
The hope was that
if there is a persistent superconducting current trapped on the string,
then  this current could in turn balance the string tension and
prevent the loop from shrinking.
Such a configuration would have a total energy
$E \sim \mu_{\mathrm{string}} L +  J^2 L$,
where $\mu_{\mathrm{string}}$ is
the string tension (or energy per unit length), $J \sim 1/L$
is the current, and
$L$ is the length of the loop. Since the energy in the global string
is linear in $L$
and the energy due to the current goes like $1/L$, there should exist a
classically stable configuration for nonzero length $L_{\mathrm{min}}$.
However, it was realized that springs
do not exist in a large region of parameter space due to the
fact that as the loop shrinks the current becomes larger, which quenches
the value of the condensate on the string \cite{turner,davisshell1}.
Physically, the current acts as a positive mass squared
in the Lagrangian, which decreases the size of the condensate in the core.
In most cases the maximum current which can occur before the condensate
is quenched is less than the current which is needed to stabilize
the string loop. Therefore, though such a spring may become a stable 
configuration (due to the special fine tuning), it is not a typical,
but rather exceptional case.
 
A more general type of topological
defect was introduced by Davis and Shellard in
\cite{davis,davisshell2,davisshell3} which has two
types of conserved  charges trapped on the string world sheet. The first
is a topological charge which had been included the analysis of springs, and
the second is a Noether charge which is also trapped on the
string world sheet. The difference with  the previous case is that a more 
general solution is ``stationary'' but not ``static;'' rather 
it has an explicit time-dependent phase $e^{i \omega t}$. This factor 
leads to a conserved charge trapped on the core, and the configuration
can  be stabilized due to the conservation of the corresponding charge. 
The time-dependent configuration becomes the lowest energy state  
in the sector with a given nonzero charge. A similar idea was
advocated by Coleman \cite{qball} in his construction of $Q$ balls, stable
objects with a time-dependent wave function. 
This class of objects generally possess nonzero angular
momentum and charge, which lead these quasiparticles to be
referred to as vortons. It is also interesting to note that numerical 
simulations performed recently in Ref. \cite{shellardnumerical} seem to 
confirm the classical stability of these vortons. 

It is the presence of a nonzero
condensate trapped on the string core which stabilizes 
these topological defects, as was shown originally in a similar model in
\cite{davis,davisshell2,davisshell3}. It was originally demonstrated
in \cite{krstrings} that there exists a nonzero condensate on the 
core of $K^0$ strings and that 
semitopological defects (known as springs in the 
cosmic string literature, as discussed above) may be present 
in high density QCD. However, we know from the cosmic string analysis 
that stable springs exist only in a very small
region of parameter space \cite{turner}. Since the effective Lagrangian
used in \cite{turner} is very similar to the one used to describe the 
CFL$+K^0$ phase of high density  we expect these results to apply in 
this case. Therefore, in this paper 
we consider the more general time dependent 
configurations (vortons)  with $\omega \neq 0$ that possess nonzero 
charge and current trapped on the string resulting in stable loops in a much 
larger region of parameter space \cite{davisshell2}.
We will apply the ideas from cosmic strings \cite{davisshell2} to high 
density QCD, extending the work of \cite{krstrings} by constructing 
vortons in the CFL$+\kzero$ phase of high density QCD. 
One difference between our case and other theories where vortons are
present is that our vortons have a domain wall \cite{sonKaon} which
stretches across the surface of the vorton like a soap bubble.
This particular type of vorton with a domain wall attached
was discussed recently within
the linear sigma model at nonzero temperature \cite{drumvorton}. 
We should
note that we will not be addressing the issue of vorton formation
in this paper. 
The physical picture is quite complicated: vortons form (due to 
angular momentum) and decay (due to weak and electromagnetic interactions).
Rather,
we will assume that there is some nonzero probability for the formation
of such topological defects when the CFL$+\kzero$ phase occurs.
We refer the reader to \cite{formation} for a detailed study on
vorton formation. 
Furthermore, there are many other issues that will not be addressed 
in the present paper. These include finite temperature effects, 
estimates on the lifetime of a vorton, weak interactions with the 
electrons present in the CFL$+K^0$ phase, and EM interactions with 
the electric and magnetic fields that are present in a neutron star.
Some of these topics (along with many useful references) can be found in 
Chaps. 5 and 8 of 
the book by Vilenkin and Shellard \cite{defectsbook} where they have been
discussed in the cosmological setting. 
We should point out that there is one new element that has 
not been discussed previously in the cosmic string literature, stability
with respect to weak interactions. This new element, along with 
stability with respect to electromagnetic interactions, will not 
be discussed in this paper and remains to be investigated. 
Our contribution in this paper is the the observation that
the formal construction suggested by Witten \cite{witten} 
and developed in many other papers 
\cite{davis,davisshell2,davisshell3,defectsbook}
may be realized in nature in the CFL$+K^0$ phase of high density QCD where 
the formal effective Lagrangian is exactly what people discussed in the works
on cosmic vortons. We hope that the present work will stimulate future 
work on the many interesting problems related to vortons in the 
CFL$+K^0$ phase mentioned above. Specifically, it would be interesting to 
understand the vorton dynamics, the rate of formation, the interaction with 
the environment (such as the electric and magnetic fields), and the 
estimation of the lifetime. 

Recently, vortons have also been constructed 
in the context of the Zhang's $SO(5)$ theory of high temperature 
superconductivity \cite{so5vorton}. This analogy between 
astrophysics or cosmology and condensed matter physics provides a unique
opportunity  to study the  cosmological or astrophysical phenomena by doing 
laboratory experiments
in condensed matter physics. Over the past few years 
several experiments have been done to test ideas drawn from cosmology 
(see the review papers \cite{kibble,volovik} for further details).

This paper is organized in the following manner. In Sec. II we will
briefly review the properties of the CFL$+\kzero$ phase of
high density QCD as well as the properties of the superconducting
$K$ string as presented in \cite{superk}. In Sec. III, we will
show that classically stable loops of superconducting $K$ strings can exist.
The domain walls which are attached to such vorton
configurations will be discussed in Sec. IV. Section IV will
also contain numerical estimates of various properties of the vortons
such as their size and the  Magnus force that leads to further stability.
Finally, we end this paper in Sec. V with concluding remarks
where we argue that the vortons can carry the angular  momentum very 
efficiently; they need the least energy per given angular momentum.

\section{The CFL+$\kzero$ phase of high
density QCD and superconducting $K$-strings}

The ground state of the high density phase of QCD
is characterized by a diquark condensate \cite{colorsc,colorsc2,CFL,rapp}
analogous to the condensate of electron Cooper pairs
present in a conventional superconductor. This phase of QCD is

referred to as a color superconducting phase and the form of the condensate
for $N_c=N_f=3$ (CFL phase) has a very specific structure
given by \cite{CFL}
\bea
\label{qqCFL}
  \<q^{ia}_{L\alpha} q^{jb}_{L\beta} \>^* &\sim&
  \epsilon_{\alpha\beta\gamma} \epsilon^{ij}\epsilon^{abc} X_c^{\gamma} ,
  \nonumber \\
  \<q^{ia}_{R\alpha} q^{jb}_{R\beta} \>^* &\sim&
  \epsilon_{\alpha\beta\gamma} \epsilon^{ij}\epsilon^{abc}Y_c^{\gamma} ,
\eea
where $L$ and $R$ represent left and right handed quarks, $\alpha$, $\beta$,
and $\gamma$
are the flavor indices, $i$ and $j$ are spinor indices, $a$, $b$, and $c$ are
color indices, and $X_c^{\gamma}$ and $Y_c^{\gamma}$
are complex color-flavor matrices describing the Goldstone bosons.
In order to describe the light degrees of freedom in a gauge invariant way,
one introduces the color singlet field $\Sigma$
\beq
\label{compsigma}
\Sigma_{\gamma}^{\beta}= X Y^{\+}  = \sum_c X_c^{\beta} Y^{c *}_{\gamma},
\eeq
One can describe the Goldstone bosons contained in the field $\Sigma$ using
the following effective Lagrangian \cite{bedascha,bedaque,schafer,kaplredd}:
\bea
\label{Leffsigma}
{\cal L}_{\mathrm{eff}}&=&\frac{\fpi^2}{4} \trace \left[
        \nabla_0\Sigma \nabla_0\Sigma^{\+}
        - \vpi^2 \d_i \Sigma \d_i \Sigma^{\+} \right], \nn \\
        &+& 2 A \left[ \det(M) \trace(M^{-1} \Sigma + h.c. \right] ,\\
\nabla_0 \Sigma &=& \d_0 \Sigma
        + i \left(\frac{M M^{\+}}{2 p_f} \right) \Sigma
        - i \Sigma \left(\frac{M^{\+} M}{2 p_f}  \right) \nn,
\eea
where the matrix $\Sigma= \exp(i \pi^a \lambda^a/ \fpi )$ describes the
octet of
Goldstone bosons with the $SU(3)$ generators $\lambda^a$ normalized as
$\trace(\lambda^a \lambda^b)=2 \delta^{ab}$.
The quark mass matrix in Eq.~(\ref{Leffsigma})
is given  by $M=\diag(m_u,m_d,m_s)$. 
The constants $\fpi, \vpi$ and
$A$ have been calculated in the leading perturbative approximation and
are given by \cite{ss,bbs,manuel,schafermass}
\beq
\fpi^2=\frac{21-8 \ln 2}{18} \frac{\mu^2}{2 \pi^2}, ~~~
\, \vpi^2=\frac{1}{3},~~~  \, A=\frac{3 \Delta^2}{4 \pi^2}.~~~
\eeq
In \cite{bedascha,bedaque,schafer,kaplredd} 
it was noticed that for a physical value of
the strange quark mass, $m_s > 60$ MeV, 
$\kzero$ condensation would occur and that $\Sigma_o=\diag(1,1,1)$
would no longer represent the true ground state of
the CFL phase \cite{bedascha}.
Instead, $\Sigma_o$ would be rotated in some different
direction in flavor space. The instability of the ground state
originates from the addition of the
covariant derivative in Eq.~(\ref{Leffsigma}).
In the following we will consider the physical
case where isospin symmetry is not exact $(m_d > m_u)$ and overall
electric charge neutrality, such that $\kzero$ condensation occurs.
The appropriate expression for $\Sigma_o$ describing the $\kzero$
condensed ground state can be parametrized as
\beq
\Sigma_o= \left( \begin{array}{ccc}
     1 & 0 & 0 \\
     0 & \cos \thetak & \sin \thetak e^{-i\varphi} \\
     0 & -\sin \thetak e^{i\varphi} & \cos \thetak
 \end{array} \right),
\eeq where $\varphi$ describes the corresponding Goldstone mode and
$\thetak$ describes the strength of the kaon condensation with
\cite{bedascha} 
\bea
\label{vevtheta} 
\cos \thetak &=& \frac{m_0^2}{\mueff^2}, ~~~a=\frac{4A}{f_{\pi}^2}
=\frac{3 \Delta^2}{\pi^2 \fpi^2},  \\
m_0^2 &\equiv& a m_u (m_d + m_s),
	~~~ \mueff\equiv\frac{m_s^2}{2p_F}. \nn
\eea 
In order for kaon
condensation $(\thetak \neq 0)$ to occur, we must have $ m_0 <
\mueff$. This leads to the breaking of the hypercharge \UY
symmetry. As discussed in \cite{bedascha}, the lightest degrees of
freedom in the CFL$+\kzero$ phase are the $\kzero$ and $\kplus$
mesons. 

Before we proceed, let us make the following  remark regarding
the description of Goldstone particles and 
topological defects
 based on the effective Lagrangian approach.
To formulate the problem in a more specific way, let us remind the reader
that, in general, the effective Lagrangian
describing the Goldstone modes can be represented in many different forms   
as long as the
symmetry properties are respected. The results for the amplitudes  
describing the interaction of the Goldstone particles do 
not depend on a specific representation used. 
A well-known example of this fact is the possibility
of describing  the $\pi$ meson properties by using
a linear $\sigma$ model as well as a nonlinear $\sigma$ model 
(and many other models which satisfy the relevant symmetry breaking pattern). 
The results  remain the same if one discusses the local properties
of the theory (such as the $\pi\pi$ scattering  length)
when the $\pi$ meson is considered as a small  quantum fluctuation rather than
a large  background field.  It may not be the case 
when $\pi$ represents a large background field in which case
some numerical difference  between different representations of the effective
Lagrangian may occur. Roughly speaking, the source of the difference is 
an inequality
$\pi(x)\neq\sin\pi(x) $ for large global background fields such as a 
string solution which is the subject of this letter.  Therefore,
in what follows we assume 
that the qualitative effects which follow from the low-energy Lagrangian
remain untouched when a different representation  for the
fields or interactions is used as long as the symmetry properties 
are respected. Our assumption is based on the experience \cite{superk}
when two different representations for $K$ fields lead to the 
similar numerical results.

With this remark in mind, we shall use the   effective Lagrangian 
by expanding the full effective Lagrangian given by Eq.~(\ref{Leffsigma}) to 
fourth order in the fields $\kzero,\kplus$. We expect that this Lagrangian
captures the essential physics of the $\kzero$ and $\kplus$
mesons because it respects all relevant symmetries:
\begin{widetext}
\bea 
\label{Leff} 
{\cal L} &=& \frac{1}{6\fpi^2} ((\bar{K}^0 \overleftrightarrow{D_{\mu}} K^0)^2 
	+ (\bar{K}^+ \overleftrightarrow{D_{\mu}} K^+)^2 
	+ (\bar{K}^+ \overleftrightarrow{D_{\mu}} K^0) 
		(\bar{K}^0 \overleftrightarrow{D_{\mu}} K^+) 
	+ (\bar{K}^+ \overleftrightarrow{D_{\mu}} K^+) 
		(\bar{K}^0 \overleftrightarrow{D_{\mu}} K^0)) \nn \\
&+& |D_{\mu}\kzero|^2 + |D_{\mu}\kplus|^2  
	- m_0^2 |\kzero|^2 - m_+^2 |\kplus|^2 
	+ \frac{m_0^2}{6 \fpi^2}|\kzero|^4 
	+ \frac{m_+^2}{6 \fpi^2} |\kplus|^4 
	+ \frac{(m_0^2 +m_+^2)}{6 \fpi^2} |\kzero|^2 |\kplus|^2 
\eea
\end{widetext}
where the covariant derivative is defined by
\bea
\label{CovDer}
D_0 &\equiv& (\d_0 - i \mueff),~~~ D_i \equiv \vpi \d_i, \nn\\
\bar{\phi}_1 \overleftrightarrow{D_{\mu}} \phi_2 &\equiv& 
	\bar{\phi}_1 (D_{\mu} \phi_2) - (\overline{D_{\mu}\phi_1 })\phi_2, \\
	(\overline{D_{0}\phi_1 }) &\equiv& (\d_0 + i \mueff) \bar{\phi}_1,~~~
	(\overline{D_{i}\phi_1 }) \equiv v \d_i \bar{\phi}_1 \nn
\eea
and the masses are given by
\beq
\label{masses} m_0^2 = a m_u (m_s+m_d), ~~~ m_+^2 = a m_d (m_s
+ m_u).
\eeq
It is necessary to keep the fourth order terms involving
derivatives such as $\d_0 \kplus, \d_i \kplus$ in Eq.~(\ref{Leff}) for a 
discussion of vortons (the reason for this will become apparent 
in the next section). 
The effective Lagrangian given in Eq.~(\ref{Leff}) will be used throughout
the rest of this paper.

If one neglects the fourth order
terms with derivatives   
$\sim \mueff \neq 0$ only,  the Lagrangian
(\ref{Leff}) reduces to one discussed in \cite{superk} expressed 
in terms of the single complex doublet $\Phi = (\kplus, \kzero)$:
\bea
\label{LsuperK} 
{\cal L} &=& |\partial_0 \Phi|^2 - \vpi^2 |\partial_i \Phi|^2   \nn \\
&-& \lambda \left( |\Phi|^2 - \frac{\eta^2}{2} \right)^2 - \delta m^2
\Phi^{\dagger} \tau_3 \Phi,
\eea
where $\tau_3$ is the third Pauli matrix and 
the various parameters are given by:
\bea
\lambda &\simeq& \frac{4 \mueff^2 - m_0^2}{6 \fpi^2},~~~ 
\lambda \eta^2 = \mueff^2 - \frac{m_0^2 + m_+^2}{2}, \nn \\
\delta m^2 &=& \frac{m_+^2 - m_0^2}{2} = \frac{a}{2} m_s (m_d - m_u) 
\eea
As we know, the Lagrangian (\ref{LsuperK}) admits superconducting
string solutions, which is not immediately obvious when $\cal{L}$ is written
in the representation (\ref{Leff}).
In everything that follows we will take $m_u \neq m_d \ll m_s$ so
that the $SU(2)$ isospin symmetry is broken. Given this, the
effective Lagrangian (\ref{Leff}) is invariant under the $U(1)_{Y}
\times U(1)_{EM}$ symmetry group, which  correspond to independent 
phase rotations for the $K^0$ and $K^+$ Goldstone bosons. When 
the explicit symmetry breaking terms are zero, the symmetry 
becomes $SU(2) \times U(1)$. 

We will briefly review the basic characteristics of the
superconducting $K$ strings as presented in \cite{superk}. If we consider
the case where $\mueff > m_0$, then the $\kzero$ field in Eq.~(\ref{Leff})
acquires a negative mass squared and this signals the formation of a
nonzero $\kzero$ condensate which spontaneously breaks the
\UY symmetry. The breaking of this \UY symmetry leads to the
existence of classically stable nontrivial solutions to the
time independent equations of motion (strings). This particular
type of topological defect is characterized by the variation of the
phase of the field $\kzero$
from $0$ to $2 \pi$ around a point where the vacuum expectation value
of $\kzero$ vanishes. Outside of this region where
the field approaches its vacuum expectation
value over a distance scale $\sim 1/m$,
where $m$ is the mass scale associated with $\kzero$.
In addition to the formation of $\kzero$ strings, in a
certain range of parameter space it is energetically favorable for
$\kplus$ condensation to occur at the center of the string as discussed
in \cite{superk}, which leads to strings that are superconducting.
The superconducting strings can be described using the following
time independent ansatz:
\bea
\label{stringsoln}
\kzero_{\mathrm{string}}(r,\phi) &=& \frac{\eta'}{\sqrt{2}} f(r) e^{i \phi}, \\
\kplus_{\mathrm{cond}}(r) &=& \frac{\sigma}{\sqrt{2}} g(r), 
\eea
where $\eta'/\sqrt{2} = \sqrt{(\mueff^2 - m_0^2)/ (2 \lambda_0)}$ is 
different from the vacuum expectation value for the field 
$\< \Phi \> = \eta/ \sqrt{2}$ [see Eq.~(\ref{LsuperK})] by  the size of 
the  symmetry breaking term $\sim \delta m^2$,
$\phi$ is the azimuthal angle in cylindrical coordinates,
$\sigma/ \sqrt{2}$ is the value of the condensate on the string
core, and $f(r)$ and $g(r)$ are solutions to the equations of
motion which obey the boundary conditions $f(0)=0,~f(\infty)=1$
and $g'(0)=0,~g(\infty)=0$. This configuration is the one
described above where the field $\kzero$ vanishes at the center of
the string and goes to its vacuum expectation value at $\infty$,
with a nonzero $\kplus$ condensate that exists only on the string
core. The functions $f(r)$ and $g(r)$ can be approximated by the
following functions which obey the appropriate boundary
conditions: 
\bea 
\label{varsoln}
f(r) &\approx& (1- e^{-\beta r}), \\
g(r) &\approx& e^{-\kappa r} (1 + \kappa r), \nn
\eea
where $\beta \simeq \sqrt{\mueff^2 - m_0^2}$
and $\kappa \simeq \delta m$
are the approximate inverse widths
of the string core and condensate respectively. In addition, the value of
the condensate at the center of the string can be estimated by substituting
the approximate solutions (\ref{stringsoln}) and (\ref{varsoln})
into the Hamiltonian  and minimizing the energy with respect to the
parameter $\sigma$.

\section{$K$ Vortons}

Now that we have reviewed the basic ideas behind superconducting $K$ strings,
we will proceed to show that superconducting string loops
can exist as classically stable objects which are supported by the presence
of two conserved charges which become trapped on the string world sheet, 
called vortons. These quasiparticles
 have been widely discussed in the context of 
cosmology \cite{witten,hindmarsh,turner,davis,davisshell1,davisshell2,
davisshell3,drumvorton}. In our case, high density QCD, we have the 
benefit of having an effective Lagrangian that contains parameters 
that have already been calculated. 

\subsection{Springs vs. vortons}

Shortly after the pioneering paper of Witten \cite{witten} on
superconconducting strings, there was a lot of interest in
the idea that superconducting strings loops
could be supported by the presence
of persistent currents \cite{hindmarsh,turner,davisshell1}.
We will consider a large loop of string of radius $R \gg \delta$, where
$\delta$ is the string thickness, so that curvature effects can be
neglected. The $z$ axis is defined along the length of the string,
varying from $0$ to $L=2 \pi R$ as one goes around the loop.
The superconducting current
can be described by including a phase in the $\kplus$ field,
$\kplus \rightarrow \kplus e^{i k z}$. Following \cite{witten}, we 
define a charge $N$ which is topologically conserved:
\beq
\label{topo}
N = \oint_C \frac{dz}{2 \pi} \left( \frac{d \alpha}{d z} \right) = k R, 
\eeq
where the path $C$ is defined along the string loop and $\alpha = k z$ 
is the phase. Since the field must be single valued
as one goes around the loop, $k$ can be interpreted as a winding number
density $k=N/R$, with $N$ constrained to be an integer. If $N$ is
an integer then it cannot change continuously.
This means that there is a persistent current associated with the
conserved quantity $N$. The only way that $N$ can unwind is through
a tunneling process whereby the condensate is quenched to down to zero
on the string, allowing the winding number
to decrease from $N$ to $N-1$ \cite{vilenkin}.

The energy of this configuration has the form
\bea
\label{springE}
E &=& \mu_{\mathrm{string}} L + v^2 k^2 L \Sigma \nn \\
	&=& \mu_{\mathrm{string}} L + (2 \pi)^2 v^2 \frac{N^2}{L} \Sigma
\eea
where $\mu_{\mathrm{string}}$ is the $\kzero$ string tension,
 winding number
density $k$ is expressed in terms of the conserved charge $N$, 
Eq.~(\ref{topo}), and the quantity $\Sigma$ is defined as the integral of
$|\kplus|^2$ over the string cross section\footnote{The quantity $\Sigma$
should not be confused with the matrix
$\Sigma^{\beta}_{\gamma}$ in the last section
that describes the octet of Goldstone bosons.}
\beq
\label{Sigma}
\Sigma \equiv \int_\times d^2 r |\kplus|^2.
\eeq
This energy has a nontrivial minimum with respect to the loop length
$L$, thus it was originally believed that springs are stable 
semitopological objects.

However, later on it was realized that the spring cannot carry 
arbitrarily large currents
or winding number densities $k=N/R$. The addition of the $z$-dependent phase
contributes an effective positive mass-squared term to the Lagrangian:
\beq
\label{nmass}
\delta {\cal L} = - v^2 |\d_z \kplus|^2 = - v^2 k^2 |\kplus|^2.
\eeq
As the loop with a conserved and nonzero charge $N$ (defined
at the moment of loop formation) shrinks to reach the energetically 
favorable length, $k$ increases, and the effective mass squared 
of $\kplus$ on the string core also
increases hence decreasing the strength of the condensate inside 
the core. Eventually it may be
no longer energetically favorable for $\kplus$ to condense 
inside the string core, and
superconductivity on the string will be destroyed with 
$|\kplus|$ quenched down to 0 on the string.
In most models discussed in the context of cosmology \cite{turner,
davisshell1}, quenching occurs before the spring reaches its
equilibrium length and hence no stable configurations exist.

However, this is not the end of the story. As Davis and Shellard
originally pointed out there exists a more general type of
topological defect which is stabilized by angular momentum
\cite{davis,davisshell2,davisshell3}. These types of objects are
referred to as vortons and have been widely discussed in the
context of cosmology and cosmic strings (see \cite{defectsbook} for a review
and \cite{shellardnumerical,shellard2002} for recent work). 
As well as the topological charge present in the 
spring configurations discussed
previously, vortons also have a conserved Noether
charge on the string core. The amount of charge present on the string is
proportional to the parameter $\omega$ in the time dependent phase
$\kplus \rightarrow \kplus e^{-i \omega t}$. The addition of this phase
leads to a nonzero Noether charge given by
\beq
\label{noetherQ}
Q = \int d^3 r j^0 = \omega L \Sigma. 
\eeq
The addition of time dependent phase also contributes
to the effective mass squared of the $\kplus$ field as in
Eq.~(\ref{nmass}), only having the opposite sign:
\beq 
\label{wmass}
\delta {\cal L} = |\d_0 \kplus|^2 = +\omega^2 |\kplus|^2. 
\eeq 
Yet $\omega$ enters the energy with the same sign as $k$: 
\bea
\label{vortonE} 
E &=& \mu_{\mathrm{string}} L + (v^2 k^2 + \omega^2) \Sigma L \nn \\
	&=& \mu_{\mathrm{string}} L +  (2 \pi)^2 v^2 \frac{N^2}{L} \Sigma +
	\frac{Q^2}{\Sigma L},
\eea
and  therefore the energy still has a nontrivial minimum with respect
to the loop length.

Thus, from Eq.~(\ref{wmass}) we see that 
a nonzero $\omega$ will counteract the quenching effect of $k$,
increasing the value of the condensate on the string
(antiquenching) as the string loop shrinks \cite{davisshell2}.
As discussed in \cite{davisshell2}, when the loop shrinks 
$\omega /(v k)$ tends to 1, meaning that at equilibrium 
length the quenching and antiquenching effects approximately
cancel each other out, leaving a stable vorton behind. 

Note that the results discussed above do not rely on whether the
condensate on the string is electrically charged. As Davis and
Shellard originally pointed out \cite{davisshell2}, the stability of the
vortons is purely mechanical and not
electromagnetic in origin. The reason is simple, a vorton 
with nonzero $N$ and $Q$ has nonzero angular momentum. 
The fact that the vorton is spinning and angular momentum is conserved
leads to the classical stability of these objects. 
Therefore,
the addition of electromagnetic effects should not change the qualitative
behavior that will be discussed below, and therefore we neglect the 
electromagnetic contribution in the present work. 

\subsection{Vortons in the CFL$+\kzero$ phase of high density QCD}

In order to describe a vorton in our case, we will add a time and $z$
dependent phase in the standard form
to the string-condensate solution
(\ref{stringsoln}) presented in the previous section:
\bea
\label{vortonsoln}
\kzero &=& \kzero_{\mathrm{string}}(r,\phi), \\
\kplus &=& \kplus_{\mathrm{cond}}(r)~e^{-i\omega t + i k z}. \nn
\eea 
Recall that $z$ is defined as the coordinate which runs along
the length of the string and varies from $0$ to $L=2 \pi R$ as one
goes around a loop of radius $R$. The loop is assumed to be large,
$R \gg \delta$ ($\delta$ is the typical string thickness), so that
we can neglect curvature effects and consider a straight string.

We can substitute these expressions into the original Lagrangian
(\ref{Leff}) and obtain the Lagrangian describing 
the dynamics in the two transverse dimensions:
\bea
\label{Klambda4}
{\cal L}&=&-\vpi^2 |\d_i \kzero|^2 -\vpi^2
|\d_i \kplus|^2 + M_0^2 |\kzero|^2 + M_+^2 |\kplus|^2 \nn \\
   &-&\lambda_0 |\kzero|^4 -\lambda_+ |\kplus|^4 - \zeta |\kzero|^2 |\kplus|^2 
\eea
where $i$ runs over $x,y$ and
\beq 
\label{coupls} 
\wt = \omega +\mueff
\eeq
is the effective frequency of $\kplus$ field, and the
parameters of Eq.~(\ref{Klambda4}) are given by
\bea
M_0^2 &=& \mueff^2 - m_0^2,~~~ M_+^2 =  \wt^2 - \vpi^2 k^2 - m_+^2 \nn \\
\label{paramwn}
\lambda_0 &=& \frac{4 \mueff^2 - m_0^2}{6\fpi^2},~~ 
	\lambda_+ =\frac{4(\wt^2 - \vpi^2 k^2) - m_+^2}{6 \fpi^2}, \nn \\
\zeta &=& \frac{(\wt +\mueff)^2 + 4 \wt \mueff - \vpi^2 n^2 
	- m_+^2 - m_0^2}{6 \fpi^2}. \nn
\eea
In simplifying Eqs.~(\ref{Leff}) to (\ref{Klambda4}) we have ignored
all fourth order terms in fields which have derivatives in $x,y$
directions since these variations change the profile (as a function of $r$)
of the string itself, but do not influence the effects that   
are the main subject of this work -- the formation of a closed loop
of the string, a vorton. 

It is important to note that fourth order couplings in
Eq.~(\ref{Klambda4}) are strongly dependent on $\omega$ and $k$. This
property is what  
distinguishes our model from the models considered in the context of
cosmology \cite{davisshell1}. However, the main feature of the
time-dependent ansatz (\ref{vortonsoln}) 
which leads to the existence of stable 
vortons does not depend on these small differences; it
remains the same as discussed earlier in different models in the cosmological 
context \cite{davisshell1}. This is because the 
stability of vortons is not related to the specific properties of the 
Lagrangian, but rather it is guaranteed by the conservation of topological 
charge (\ref{topo}) and the Noether charge: 
\bea
\label{noether}
Q &=& \int d^3 r j^0 \nn \\
	&=& \frac{L}{2} \int d^2 r ~|\kplus|^2 
	\left[ 2(\omega + \mueff) \right]  \nn \\ 
	&\simeq&  L \wt \Sigma 
\eea
which reduces to the expression (\ref{noetherQ}) with the replacement
$\omega \rightarrow \wt = \omega + \mueff$.
In the above expression we have omitted higher order terms\footnote{As we 
have already mentioned these higher order terms in the effective description 
lead, in general,  to some difference in the definition of the fields. 
In particular, we could define the fundamental field $K^+$ as the phase of 
$\Sigma^{\beta}_{\gamma}$ (\ref{compsigma}) 
or we could define $K^+$ as $\sin (K^+)$ etc. We do 
not expect that this ambiguity can change the qualitative results
which follow.} in derivatives and/or fields to simplify the 
expression for the charge $Q$ (\ref{noether}).
The presence of
time dependence leads to a configuration with nonzero
conserved Noether charge. Such an object, which is  stable 
due to the conservation of a Noether charge,
is called a $Q$ ball \cite{qball}. Vortons have been referred
to as semitopological defects \cite{davis} due to the 
fact that they are partially stabilized by
topology and partially stabilized by the presence of a 
conserved Noether charge, similar to $Q$ balls. 

Thus $\kplus/\kzero$ 
string loops in our model are always charged and this charge has
to be taken into account when studying their dynamics. In
particular, even if the $\kplus$ field originally has no explicit
time dependence, i.e., $\omega = 0$ and $\wt = \mueff$, explicit
time dependence will appear in the process of loop shrinking to
preserve the Noether charge that was present at the moment of formation.

Now we proceed to study what values $\wt$ and $k$ can assume.
These values are clearly not arbitrary since the masses and the
couplings in the Lagrangian (\ref{Klambda4}) depend on $\wt$ and $k$.
Thus, for the vorton to exist, $\wt$ and $k$ must not destroy 
the $K^+$ condensate of the  superconducting $\kzero/\kplus$ string. The
constraints on the parameters in Eq.~(\ref{Klambda4}) which guarantee
superconductivity have been discussed in detail by \cite{turner}
and can be stated as follows:
\begin{enumerate}
\item{It must be energetically favorable for $\kzero$ to condense in
the vacuum, i.e. in vacuum $\<\kzero\> \neq 0$. This guarantees that
the $U(1)_{Y}$ symmetry is spontaneously broken and a $\kzero$
string can form.}
\item{It must be energetically unfavorable for $\kplus$ to condense
in the vacuum, i.e. in vacuum $\<\kplus\> = 0$. This guarantees that
$\kplus$ does not condense outside the string and is bound to the
string core. This constraint requires that the effective
mass squared of $\kplus$ must be positive off of the string core.}
\item{It must be energetically favorable for $\kplus$ condensation 
to occur on the string core. 
A necessary condition for this is that the
effective mass squared of $\kplus$ must be negative inside the string core.}
\item{A sufficient condition for 3. is that total energy associated
with $\kplus$ condensation inside the string core must be
negative.}
\end{enumerate}
The first 3 constraints can be summarized in terms of the parameters 
(\ref{paramwn}) as
\bea
1&.& \frac{M_0^4}{\lambda_0} >
\frac{M_+^4}{\lambda_+} \nn \\
\label{constraints}
2&.& \frac{\zeta}{2 \lambda_0} M_0^2 > M_+^2 \\
3&.& M_+^2 > 0 \nn
\eea
Before we make numerical estimates for the parameters $\omega,k$, notice
that the approximate degeneracy between $K^0$ and $K^+$ implies
that $M_+^2 \simeq M_0^2$ and therefore
\beq
\label{chiral} 
\wt^2 - \vpi^2 k^2 - m_+^2 \approx \mueff^2 -m_0^2, 
\eeq 
which suggests that the typical scale for $\omega$ and $k$ is the 
scale $\mueff \sim m_s^2/(2 \mu) \sim 30$ MeV. Our variational 
calculations [similar to the one performed in \cite{superk} 
using the ansatz (\ref{varsoln})]
used to obtain the allowed values of $\omega$ and $k$ support this estimate. 
When values of $\wt$ and $k$ satisfying all 
4 conditions are plotted 
we see that superconducting strings exist for $\wt$
from 25 to 45 MeV. 

The upper limit on $\wt$ has to do with the fact that for large
values of $\wt$ the parameters $M_+^2$ and
$\zeta$ break the degeneracy
between the $\kzero$ and $\kplus$ fields. Moreover, the increase of
$M_+^2$ and $\zeta$ with $\wt$ cannot be cancelled out simultaneously
by an increase in $k$. Thus for large $\wt$, $\zeta$ becomes too large
and the energy associated with a $\kplus$ condensate in the 
core is no longer negative.

The total energy of this field configuration can also be computed
from Eq.~(\ref{Leff}). The energy is given as usual by the integral of
the $T^{00}$ component of the energy-momentum tensor: 
\bea
\label{energy} 
E &=& L\int d^2 r (\vpi^2 |\d_i \kzero|^2 +
\vpi^2 |\d_i \kplus|^2 - M_0^2 |\kzero|^2   \\
&-& M_+'^2 |\kplus|^2+\lambda_0 |\kzero|^4 + \lambda'_+ |\kplus|^4 
	+ \zeta' |\kzero|^2 |\kplus|^2) \nn
\eea
where the coefficients are given by
\bea 
\label{energy_coeff}
M_0^2 &=& \mueff^2 - m_0^2,~~
	\lambda_0 = \frac{4 \mueff^2 - m_0^2}{6\fpi^2}, \nn \\
M_+'^2 &=& \mueff^2 - \omega^2 - v^2 k^2 - m_+^2, \nn \\
\lambda'_+ &=&
	\frac{4 (\mueff^2 - \omega^2 - v^2 k^2) - m_+^2}{6 \fpi^2}, \\	
\zeta' &=& \frac{8 \mueff^2 - \omega^2 
	- v^2 k^2 - m_+^2 - m_0^2}{6 \fpi^2}. \nn
\eea
In the case when $\omega = k =0$ one reproduces the energy for the
string obtained from Eq. (\ref{LsuperK}). 
The part of the energy (\ref{energy})
associated with a single $\kzero$ vortex of length $L$ without a
condensate in the center [terms involving only $\kzero$ in Eq.~(\ref{energy})]
is given to logarithmic accuracy as 
\beq 
\label{Ek0} 
E_{\kzero} = 2 \pi R ~(\pi \eta'^2 v^2 \ln(\beta \Lambda)) ,~~~ 
	\eta'^2 = \frac{\mueff^2 - m_0^2}{\lambda_0} 
\eeq 
where $\beta$ is the inverse width of the string's core introduced in
the ansatz (\ref{varsoln}) and $\Lambda$ is a long distance cutoff which is
introduced in order to control the logarithmic divergence which
appears due to the large distance variation of the phase. The
cutoff is typically the distance between strings or the radius of
curvature, and since we will be considering loops of strings in
this paper, the natural correspondence to make is $\Lambda \simeq
R$. The additional energy of the $\kplus$ condensate in the core of the 
string (due to the nonzero values of 
$\omega , k$) has the leading term behavior:
\bea
\label{Ekp} 
E_{\kplus} &\simeq& L (\omega^2 + v^2 k^2) \Sigma \nn \\
	&\simeq& \frac{Q^2}{2 \pi R \Sigma} + \frac{2 \pi v^2 N^2 \Sigma}{R},
\eea 
where we expressed $\omega , k$ in terms of the conserved charges $Q , N$.
Note this is only an approximate expression for the energy of the 
condensate and that we have neglected various 
higher order terms in  $K^0,K^+$ in Eq. (\ref{energy}) as well as in 
the definition of the charge $Q$ (\ref{noether}). As we mentioned earlier, 
these higher order terms reflect the ambiguity in the description 
of solitons using the effective Lagrangian approach when $K^0/\fpi \sim 1$.
These terms effectively  play  a role by determining the magnitude of the 
$K^+$ condensate in the core represented by the parameter $\Sigma$ in 
our calculations.  Once the presence of a $K^+$ condensate  is established, 
these terms can change some numerical results, but we do not expect
that these terms can change our qualitative results because the 
existence of vortons is based on conservation of charges rather than
on the specific properties of the field representations used in this paper. 
In other words, once the parameters are such that a $K^+$ condensate forms
in the core of the vortex, the vorton can also form.
We use the simplest possible expressions for the relevant parameters 
in order to illuminate the fact that stability occurs for a nonzero
value of $R$. 
A simple intuitive explanation for the $1/R$ dependence in Eq.~(\ref{Ekp})
is as follows. 
If  we consider a fixed value of $Q$, shrinking the loop will result 
in an increased $K^+$ density on the loop. This in turn increases  the 
energy, which eventually cancels the string  tension. In the same manner, 
the momentum increases for a fixed value of $N$ for decreasing $R$, which
also counteracts the string tension. 
Thus, as discussed before, the total energy $E =
E_{\kzero} + E_{\kplus}$ has a minimum with respect to $R$ at which
a stable vorton exists. With our parameters it happens at $R_0$ given by
\beq
\label{R_0}
(2\pi R_0)^2\simeq \frac{Q^2 +  (2 \pi)^2 v^2 N^2 \Sigma^2 }
{\pi \Sigma\eta'^2 v^2 \ln(\beta \Lambda )}.
\eeq
At this point in our discussion the size of vortons is not constrained 
in any way; it could be arbitrarily
large, similar to the cosmic string vortons. However, when an explicit 
symmetry 
breaking term is taken into account, the vorton size can not be arbitrarily 
large. Rather, the size will be constrained by the strength of an  additional 
force due to the domain wall attached to the string (see the next section).

One should emphasize at this point that 
the source of this stability is purely mechanical, and not related to the 
electromagnetic interactions. This is in contrast with the 
suggestion made in \cite{krstrings}, where it was mentioned  
that it may be possible to have 
classically stable $K$ vortons in high density QCD
due to a persistent superconducting current trapped on the string.
As we have demonstrated above, the source of the vorton  stability has 
a quite different origin.
We expect that the maximum electromagnetic current which can occur 
in the system
(before the $K^+$ condensate is quenched) is less than the current 
which is required
to stabilize the string loop, as it was demonstrated to happen in most
cases \cite{turner} and \cite{davisshell1}  where
a similar problem was previously analyzed. 
Since it was demonstrated in \cite{turner} that stable springs 
are only possible in very limited region of parameter space for similar
$\phi^4$-type models, we do not want to repeat 
these calculations in the present work because our effective Lagrangian
is essentially the same and we expect that these results would apply. 

The stability of the vortons can also be demonstrated explicitly 
in a different way. As Davis and Shellard originally pointed out, 
the source of this stability is purely mechanical. The presence of time
dependence in Eq.~(\ref{vortonsoln}) allows the vortons 
to spin and carry angular momentum. The conservation of angular momentum
is reflected by the conservation of the topological and Noether charges
$Q$ and $N$, respectively.
We can easily calculate the approximate 
angular momentum carried by a vorton from the energy-momentum tensor 
obtained from Eq.~(\ref{Leff})
\beq
\label{energymomentum}
M^{ij} = \int d^3 r (x^i T^{0j} - x^j T^{0i}).
\eeq
The angular momentum carried by a single vorton (\ref{vortonsoln}) is
approximately
\beq
\label{angmom}
M \simeq 2 \pi R^2 k ~ \wt \Sigma . 
\eeq
The direction of $M$ is perpendicular to the surface formed by the vorton. 
From the expression (\ref{angmom}) for the angular momentum, we can see that
that $M \simeq N \cdot Q$ is proportional to the classically conserved 
quantities $N,Q$. 
The presence of nonzero $\omega$ is the only way to have a nonzero 
component of the energy-momentum tensor $T^{0 i} \sim \d_0 K^+$, 
which gives a nonzero
contribution to the angular momentum tensor (\ref{energymomentum}). We 
refer the reader to \cite{magnus} for further details on the relationship 
between nonrelativistic vortices and relativistic strings in a nontrivial
background. In this paper \cite{magnus} it was 
demonstrated that a correspondence can be
made between the two systems when the relativistic strings are 
put into a medium, resulting in nontrivial time dependence. This 
nontrivial time dependence yields
a nonzero magnitude for $T^{0 i}$ and the angular momentum $M^{ij}$ 
correspondingly. Therefore, nonzero charge $Q$ trapped on the vortex 
implies that a vorton carries nonzero angular momentum.

Although our $\kplus$ field is electrically charged, we have not
mentioned or included interactions with electromagnetic gauge
field. We expect that the quantitative results discussed above
would be slightly different upon including a gauge field, with the
qualitative behavior remaining unchanged. 
Qualitatively, we expect that the electromagnetic interactions
would enhance that stability of the vortons because the electromagnetic charge
of the $K^+$ condensate trapped in the core gives an additional
contribution $\sim Q^2/R$ and prevents the vortons from shrinking. 


\section{Domain walls, drum vortons, and Magnus forces}

In the previous section we have demonstrated that loops of 
superconducting $K$ strings, called vortons, can exist as classically 
stable objects due to the fact that charges and currents are trapped
on the string core. We  will now include a brief discussion 
of other effects that are important in order to have a correct 
description of vortons in the CFL$+K^0$ phase of high density QCD. 

Up to this point, we have not included terms 
in the Lagrangian that explicitly break the $U(1)_Y$ symmetry. 
If the weak interactions are
taken into account, there is a small piece which
must be added to the effective Lagrangian (\ref{Leff}) which
explicitly breaks the \UY symmetry \cite{sonKaon}:
\beq
\label{dwpotential}
\delta {\cal L} = -V(\varphi) = \fpi^2 m_{\mathrm{dw}}^2 \cos \varphi, 
\eeq
where $\varphi$ is the phase of $K^0$ and $m_{\mathrm{dw}}$ 
will be given below. As described in full detail in
\cite{sonKaon}, this leads to the formation of domain walls, with the 
phase $\varphi$ varying from $0$ to $2 \pi$ across the wall (the same 
vacuum state exists on both sides of the wall). Consequently, this 
leads a domain wall being attached to every string. 
Therefore, as one encircles the string
at large distances from the core the
phase variation from $0$ to $2 \pi$ is not uniform but is sandwiched
inside a domain wall of width $\mdw^{-1}$, 
with $\mdw^{-1}$ set by the coefficient of the
explicit symmetry breaking term. We will simply state the results of
Son \cite{sonKaon} here without going into details.
The domain wall tension $\alpha_{\mathrm{dw}}$
(energy per unit area) is given by
\bea
\alpha_{ \mathrm{dw}} &=& 8 v \fpi^2 m_{\mathrm{dw}}, \\
	m_{\mathrm{dw}}^2 &=& \frac{162 \sqrt{2} \pi}{21 - 8 \ln 2} 
	\frac{G_F}{\alpha_s} \cos \theta_c \sin \theta_c m_u m_s \Delta^2, \nn
\eea
where $G_F$ is the Fermi constant, $\theta_c$ is the Cabibbo angle, and 
$\alpha_s$ is the strong coupling constant. 
The inverse mass $\mdw^{-1}$ is approximately the width of the
domain wall. Son calculates $m_{\mathrm{dw}} \sim 50$ keV
for physical values of
the relevant parameters \cite{sonKaon}. If the size of the domain wall
is greater than the thickness, $R \gg \mdw^{-1}$ for a circular
domain wall of radius $R$, then the total energy of this configuration can
be approximated as
\beq
\label{Edw}
E_{\mathrm{dw}} \simeq  \pi R^2 \alpha_{\mathrm{dw}}.
\eeq
Since every string must be attached to a domain wall,
the vortons discussed in the previous section will have a domain wall
stretched across their surface like a soap bubble.
Similar configurations have been recently
studied in the linear sigma model at nonzero temperature
\cite{drumvorton} and have been referred to
as ``drum vortons.'' In the case that
there is no domain wall attached to the vorton, the minimization
of the energy with respect to $R$ leads to the result that
$k = N/R = {\mathrm{const}}$ for the chiral case
$v k = \omega$, independent of $R$. Therefore, the vorton
could have an arbitrarily large size. The presence of
the domain wall will lead to an upper bound on the radius of these
vortons. Now that we have an approximate expression to the domain wall
contribution to the energy, we can add
Eqs. (\ref{Ek0}), (\ref{Ekp}), and (\ref{Edw}) to arrive at an expression
for the total energy of a circular drum vorton of radius $R$
which is valid for $R \gg m^{-1}_{\mathrm{dw}}$:
\bea
\label{Etotal}
E &=& E_{\kzero} + E_{\kplus} + E_{\mathrm{dw}} \nn \\
  &=& 2 \pi^2 R~ \eta^2 v^2 \ln(\beta R)
    + \pi R^2 \alpha_{\mathrm{dw}} \nn \\
    &+& \frac{1}{R} \left(\frac{Q^2}{2 \pi \Sigma} 
	+ 2 \pi v^2 N^2 \Sigma \right). 
\eea
This expression must be minimized with respect to $R$
to find the size of these classically stable objects. 
This problem is quite complicated because of a number
simplifications we have made in Eq. (\ref{Etotal}). In particular, 
Eq. (\ref{Edw}) is not literally valid  for relatively small 
$R_0\sim \mdw^{-1}$ when equilibrium is reached (see below).  
 
We will now estimate the typical size of a vorton.
We start with relatively small charges $Q, N$ (and correspondingly $R$) 
when the domain wall contribution can be neglected,
and the equilibrium is reached at $R_0$ given by Eq. (\ref{R_0}). 
We slowly increase $Q$ and $N$
such that domain wall contribution becomes of the same order of magnitude 
as the string-related terms.
This happens when $Q, N \sim f_{\pi}/ \mdw \gg 1$. The size of the 
configuration at this point
 $R_0\sim Q/f_{\pi}\sim \mdw^{-1}$ reaches   the magnitude of the 
domain wall width,
i.e. $(50~\mathrm{keV})^{-1} $ which is much larger than any QCD-related scale 
of the problem.
If one increases $Q$ and $N$ further, the first term in Eq.~(\ref{Etotal}) 
becomes irrelevant, 
and equilibrium is achieved when 
$R_0^3 \sim Q^2/(\Sigma\alpha_{\mathrm{dw}})$ 
at which point the energy of
configuration $E\sim Q^{4/3}\alpha^{1/3}$ grows too fast with $Q$. Such 
a large configuration will decay to smaller vortons by preserving 
the charges $Q$ and $N$, decreasing the 
total energy. Therefore, one expects that the maximum vorton size 
is related to the weak interactions which  set the typical vorton 
scale to be $\mdw^{-1}$.

There exists an additional force which may further stabilize the
vortons. This is the Magnus force, which arises when a global
string moves through a Lorentz-noninvariant fluid. We naturally
have such a background, since we are working at nonzero chemical potential,
which breaks Lorentz invariance. 
The corresponding expression has been derived in Ref.~\cite{magnus} where it
was demonstrated that in the language of the Goldstone boson such a background
corresponds to a time dependent phase of the order parameter.
This phase in our notations takes the form $\sim e^{i\mueff t}$.  
If vorton moves with velocity
$\vec{v}$ through this fluid, the force exerted on the vorton per unit length:
\beq
\label{magnus}
\vec{F} = 2\pi\gamma \eta^2\mueff~ \vec{v} \times \vec{m},
\eeq
where $\gamma$ is the standard relativistic factor and
$\vec{m}$ is the circulation vector of unit magnitude, $|\vec{m}| = 1$,
which points in the direction of the string.
If the velocity vector $\vec{v}$ is perpendicular to the plane formed by
the vorton, then there will be a Magnus force present which points
outward, further stabilizing the vorton. This will in turn increase the 
size of the vorton. If the vorton moves in 
the opposite direction, the Magnus force points inward, which decreases 
the size of the vorton.

Finally, the issue of quantum stability of vortons has not been addressed
in this paper. In the pure
current case ($Q=0, N \neq 0$), the instanton solution has been explicitly
constructed and the lifetime calculated analytically \cite{vilenkin}.
The decay mechanism is a quantum mechanical tunneling process where the 
condensate goes to zero on the core, allowing the winding number
to decrease from $N$ to $N-1$. 
However, the vortons discussed here have nonzero $Q,N$ so the results
obtained in \cite{vilenkin} do not apply. In spite 
of this fact, we expect that the vortons in high density QCD 
discussed in this paper are long lived due to the approximate
``chiral'' relation (\ref{chiral}) which must be satisfied in order 
to have superconducting strings. 
We expect that the decay rate is exponentially 
suppressed as the decay of a vorton is associated with tunneling processes. 
These tunneling processes may be due to weak interaction processes, 
among other things. In general, we expect that the lifetime is 
relatively long lived because of the fact that the
tunneling decay rate should be quite small for such a large object. 
However, at the moment we cannot make 
any definitive
statements on the lifetime of the vortons discussed in this paper. 
In order to make such estimations we need to understand the vorton 
interactions, which were completely ignored in this work. 
To understand the dynamics of vortons we need to 
know: first of all, the interaction of the Goldstone particles 
with the vorton. This would allow us to calculate
the corresponding cross section which is important for 
the analysis of the frictional force
acting on a moving vorton. Secondly, the same interaction would 
allow us to estimate the Goldstone mode production by the vorton. 
This knowledge is essential for the study of the Goldstone boson radiation 
from moving  vortons. Finally, the interaction is 
essential for studying such issues 
as the typical lifetime of a vorton, the typical 
behavior  of vortons when they can 
join or disjoin with each other and absorb or emit the Goldstone
bosons. The quantum numbers ($N, Q, M$) should be conserved in 
all the processes mentioned above.
Unfortunately, none of these questions can be answered at this point.


\section{Conclusion}

In this paper we have shown that loops of superconducting $K$ strings
\cite{krstrings,superk}, called vortons,  can exist as classically
stable objects within the CFL$+\kzero$ phase of high density QCD.
These vortons are certainly topological trivial configurations as 
was explained in the original papers (see Ref. \cite{defectsbook}). 
However, if the configuration is very large in size, it might have
a very large lifetime. We have  not presented specific 
estimates for the lifetime in our case, but we hope that it 
is the same  as in previous studies and quite large \cite{vilenkin}. 
The main mechanism which stabilizes these superconducting $K$-string
loops is the presence of charge and current which is trapped
on the string. The main difference between these vortons and
vortons within other models is the presence of a domain wall which
is stretched across the  surface. These domain walls set up upper bound
on the allowable vorton size which is $\mdw^{-1}$ in contrast with 
cosmic vortons which could become  arbitrary large in size.

The most intriguing aspect of these vortons is their ability to carry
angular momentum due to the presence of nonzero charge and current trapped 
in the core. Moreover, the vortons are very efficient carriers of 
the angular momentum. Indeed, to simplify our estimates in what follows,
we use $\Lambda_{QCD}$ to be a typical scale of the 
problem, it could be any of dimensional parameters 
(or their combination) discussed above such as
$\mu_{eff}, \Delta, k, \omega, f_{\pi}$, etc. 
As we demonstrated  above, the angular
momentum carried by a single vorton of size $L$ is 
$M\sim \Lambda_{QCD}^2L^2$ [see Eq. (\ref{angmom})],
 and grows proportional to the 
area $L^2$ up to a maximal 
possible size, which is $\bar{L}\sim \mdw^{-1}$.
As we mentioned, the parameter $\mdw$ has a characteristic scale 
of the weak interactions,  
and it is three order of magnitudes smaller than $\Lambda_{QCD}$.
At the same time, the energy of the vorton scales linearly with the size, 
$E\sim \Lambda_{QCD}^2 L$
as long as $L\leq \bar{L}\sim \mdw^{-1}$. Therefore, the angular momentum 
per energy scales as
$M/E\sim L$ for $L\leq \bar{L}\sim \mdw^{-1}$. At the same time, a 
typical straight vortex (not a vorton) is expected to   carry the
angular momentum according to the relation  
$M \sim \Lambda_{QCD}L$ and  $E \sim \Lambda_{QCD}^2L$, such that 
$M/E\sim \Lambda_{QCD}^{-1}$. Therefore, the vortons are much more 
efficient carriers of the angular momentum than any regular straight vortices. 
In addition to this, the larger the vorton, the more efficient they become
at carrying angular momentum. 
However, as explained above there is a maximum vorton size
before they become unstable; it is $\bar{L} \sim \mdw^{-1}$. 

Therefore, one should expect that
most of the vortons in the core of a neutron
star would have one and the same typical size, which is 
$\bar{L} \sim \mdw^{-1}$.
As  discussed above, the vortons ability to carry
angular momentum efficiently makes them the important dynamical 
degrees of freedom. In particular, they might
be the key  elements for the explanation of phenomenon such as 
glitches. The same vortons might be important objects for other problems such 
as describing
the dynamics of the electromagnetic fields in cores of the neutron stars 
(vortons are positively charged
configurations due to a $K^+$ condensate trapped in the vortex).
The vortons could be important for discussions of transport properties,
as well as problems related to the cooling of the system. This is due 
to the fact that a vorton is a relatively large configuration with fields 
correlated over large distances (in QCD units). In such a case, as is known, 
the cross section for the particle scattering by strings,
could be very large, and could influence the cooling of the system.
Finally, the vortons
might be the only possible  carriers of the angular momentum in
 the crystalline superconducting phase \cite{krishna}. Indeed, in this phase
it is quite difficult (if possible at all) to construct a regular straight 
vortex which can carry the angular momentum.
To conclude: we believe that the physics of vortons 
could prove to be interesting for compact 
astrophysical objects such as neutron stars where CFL phase is likely 
realized. We hope that the present paper will initiate some activity 
in the direction of the phenomenological implications of vortons. 

Furthermore, we believe that the study of such objects is important 
due to a  completely different reason. Vortons were originally introduced 
in the context of cosmology 
\cite{davis,davisshell2,davisshell3} and more recently within
the $SO(5)$ theory of high $T_c$ superconductivity \cite{so5vorton}. 
In this paper we argued that vortons may play an  important role in 
astrophysics. The analogy between all these fields
provides another example where astrophysical and cosmological 
phenomena have similarities with systems
in condensed matter physics, and therefore, may be studied by doing 
laboratory experiments. For further details on recent 
condensed matter experiments designed in order to test ideas drawn from 
cosmology we refer the reader to the review papers \cite{kibble,volovik}.

\section*{Acknowledgments }

We are grateful to Robert Brandenberger for useful discussions.
One of us (A.R.Z.) is thankful to KITP, Santa Barbara, for the organization of
the workshop ``QCD And Gauge Theory Dynamics In The RHIC Era", 
which initiated this study. 
This work was supported in part by the Natural Sciences and Engineering
Research Council of Canada.

\end{document}